\documentclass[12pt]{iopart}
\usepackage{graphicx}

\begin{document}
\title[Boundary correlators of Ising model and Toeplitz system]{Boundary correlation function of fixed-to-free bcc operators in square-lattice Ising model}
\author{Seung-Yeop Lee}
\address{James Franck Institute, University of Chicago, 5640 South Ellis Avenue, Chicago, Illinois 60637, USA}
\ead{duxlee@uchicago.edu}

\newcommand{\nn}{\nonumber}
\newcommand{\bea}{\begin{eqnarray}}
\newcommand{\eea}{\end{eqnarray}}
\newcommand{\beq}{\begin{eqnarray}}
\newcommand{\eeq}{\end{eqnarray}}
\newcommand{\sh}{{\rm sh}}
\newcommand{\ch}{{\rm ch}}
\newcommand{\s}{\sigma}
\newcommand{\K}{\overline K} 
\newcommand{\D}{\theta}      
\newcommand{\V}{K}           
\newcommand{\rmf}{{\rm f}}
\newcommand{\rmb}{{\rm b}}

\begin{abstract}
We calculate the boundary correlation function of fixed-to-free boundary condition changing operators in the square-lattice Ising model.  
The correlation function is expressed in four different ways using $2\times2$ block Toeplitz determinants.  We show that these can be transformed into a scalar Toeplitz determinant when the size of the matrix is even.
 To know the asymptotic behavior of the correlation function at large distance we calculate the asymptotic behavior of this scalar Toeplitz determinant using the Szeg\"o's theorem and the Fisher-Hartwig theorem.
 At the critical temperature we confirm the power-law behavior of the correlation function predicted by conformal field theory.  
\end{abstract}

\section{History and outline}
Onsager's solution \cite{Onsager} for the two dimensional Ising model proved to be an important stepping stone for the development of a theory of critical phenomena, and for the invention of conformal field theory (CFT).  The solution has been studied by various methods: by using essentially the transfer matrix technique and the Jordan-Wigner transformation  \cite{Kaufman,Mattis,Kadanoff,Abraham}, by taking a combinatorial approach of using the Dimer problem \cite{MontrollReview,Kasteleyn,Wu}, by using Grassman variables \cite{Plechko}, and so on.  These methods lead to many exact results including the evaluation of the spin-spin correlation function \cite{Kadanoff,Wu,Montroll}, the boundary magnetization and the boundary spin-spin correlation functions \cite{Wu}, the interface free energy and the magnetization profile across the interface \cite{Abraham}, and so on. 

At the critical temperature many of these results are predicted by conformal field theory (CFT).  The rational CFT \cite{BPZ} of the central charge $c=\frac{1}{2}$ describes the Ising system at the critical temperature, producing correct operator contents and correlation functions \cite{CFTising}.  It also describes the conformally invariant boundary conditions and predicts the dimensions of boundary operators \cite{Cardy}.

In this paper we will calculate the correlation function of fixed-to-free boundary-condition-changing (bcc) operator directly from the lattice model.
By doing so we will confirm one of the predictions of CFT, namely, the dimension of fixed-to-free bcc operator, and also obtain the off-critical behavior of the correlation function.   We use the formulation by Kadanoff \cite{Kadanoff} to write the general boundary correlation functions as a $2\times2$ block Toeplitz determinant.  The asymptotic behavior of the Toeplitz determinant when the size of the matrix goes to infinity is, in many cases, given by the Szeg\"o's theorem \cite{WidomBlock}. At the critical temperature, however, the Szeg\"o's theorem does not capture the asymptotic behavior of the determinants and we must use the Fisher-Hartwig theorem to deal with the singularity.
This is possible since we can transform our block Toeplitz determinant into a scalar Toeplitz determinant when the size of the matrix is even. 

The basic definitions and theorems about Toeplitz system are collected in Appendix.

\section{Boundary correlation function: General}
\subsection{Partition function}
We consider a 2-dimensional Ising model on a square lattice.
Lattice sites are labeled by two integers $(j,k)$ representing $x$ and $y$ coordinates. 
The spin variable at $(j,k)$ is denoted by $s_{j,k}=\pm1$.
The partition function of the Ising system is given by
\begin{eqnarray}\label{first}
Z=\sum_{\{s\}}\exp\left[\sum_{j,k=-\infty}^\infty(\K_{j,k}s_{j,k}s_{j+1,k}+\V_{j,k}s_{j,k}s_{j,k+1})\right],
\end{eqnarray}
where the {\it horizontal coupling}, $\K_{j,k}$, is assigned to the bond connecting $(j,k)$ and $(j+1,k)$ and the {\it vertical coupling}, $\V_{j,k}$, is assigned to the bond connecting $(j,k)$ and $(j,k+1)$.  
For the horizontal coupling, $\K_{j,k}$, we also define the dual coupling, $\theta_{j,k}$, such that $\tanh\D_{j,k}=\exp({-2\K_{j,k}})$.

It is derived in \cite{Kadanoff} that the above quantity is evaluated as
\begin{eqnarray}\label{partition function}
 Z=\left(\prod_{j,k}2\sinh2\K_{j,k}\right)^{\frac{1}{2}}\,\sqrt{{\rm det}\big[{\sf Q}-\rme^{-\rmi p_x}\big]},
\end{eqnarray}
where ${\sf Q}$ and $\rme^{-\rmi p_x}$ are infinite matrices that we describe next.

These matrices have a block structure: the components of the matrix are labeled by indices $j,k$ and $\tau$ where the last one takes either $1$ or $2$.\footnote{This reflects the two types of fermions in the model \cite{Kadanoff}.}
First we define the {\it translation matrices} as
\begin{eqnarray}\nn
\eqalign{[\rme^{\pm\rmi p_x}]_{j,k,\tau,j',k',\tau'}=\delta_{k,k'}\delta_{j\pm 1,j'}\delta_{\tau,\tau'},\\[\rme^{\pm\rmi p_y}]_{j,k,\tau,j',k',\tau'}=\delta_{k\pm 1,k'}\delta_{j,j'}\delta_{\tau,\tau'}.}
\end{eqnarray}
These simply raise or lower the corresponding indices by one.  
We also define {\it partial-translation matrices}, ${\rm T}_\pm$, such that they raise or lower the composite index ($2k+\tau$) by one. Using $2\times2$ notation to explicitly show their action on $\tau$-space, they are written as ${\rm T}_\pm={\sigma}_1\exp(\pm\rmi \case{p_y}{2})\exp(\rmi\case{p_y}{2}{\sigma}_3)$ or, equivalently, as
\begin{eqnarray}\nn
{\rm T}_-=\left(\begin{array}{cc}0&\rme^{-\rmi p_y}\\1&0\end{array}\right)\qquad
{\rm T}_+=\left(\begin{array}{cc}0&1\\\rme^{\rmi p_y}&0\end{array}\right),
\end{eqnarray}
using the standard Pauli matrices,
\begin{eqnarray}\nn
\sigma_1=\left(\begin{array}{cc}0&1\\1&0\end{array}\right)\qquad
\sigma_2=\left(\begin{array}{cc}0&-\rmi\\\rmi&0\end{array}\right)\qquad
\sigma_3=\left(\begin{array}{cc}1&0\\0&-1\end{array}\right).
\end{eqnarray}
Using these (partial-)translation matrices we now define 
the ${\sf Q}$-matrix as
\begin{eqnarray}\label{Q}
{\sf Q}={\rm T}_+\, \rme^{-2{\sf\Theta}\sigma_2}\,{\rm T}_-\,  \rme^{\rmi p_x} \rme^{-2{\sf K}\sigma_2}\rme^{-\rmi p_x},
\end{eqnarray}
where ${\sf\Theta}$ and ${\sf K}$ are diagonal matrices defined by the (dual) couplings as
\begin{eqnarray}\nn
{\sf\Theta}_{j,k,\tau,j',k',\tau'}&=&
\delta_{j,j'}\delta_{k,k'}\delta_{\tau,\tau'}\,\D_{k,j}
\\\nn
{\sf K}_{j,k,\tau,j',k',\tau'}&=&
\delta_{j,j'}\delta_{k,k'}\delta_{\tau,\tau'}\V_{k,j}\,.
\end{eqnarray}
The partition function (\ref{partition function}) is invariant under the following similarity transformation.
\begin{eqnarray}\label{Qtilde}
{\sf Q}\rightarrow\tilde{\sf Q}=\rme^{-2{\sf \Theta}\sigma_2}\,{\rm T}_-\,  \rme^{\rmi p_x} \rme^{-2{\sf K}\sigma_2}\rme^{-\rmi p_x}\,{\rm T}_+\,.
\end{eqnarray}

\subsection{Boundary correlation function}

Using (\ref{partition function}) we will derive the expectation value of two boundary operators separated by $n$ bond-lengths.
First, following \cite{Kadanoff}, let us briefly describe how the expectation value of an operator in the bulk may be evaluated.

An operator ${\cal O}$ may be realized on the lattice by changing the couplings in a specific way (see, for instance, \cite{Kadanoff2}).
In such case the insertion of ${\cal O}$ will modify ${\sf Q}$, say, to ${\sf Q}'$. Then, $\langle{\cal O}\rangle$ may be written using (\ref{partition function}) as
\begin{eqnarray}\label{partitionbulk}
\langle{\cal O}\rangle\sim\frac{Z[{\sf Q}']}{Z[{\sf Q}]}=\sqrt{\left(\prod_{j,k}
    \frac{\sinh2\K'_{j,k}}{\sinh2\K_{j,k}}\right)\det\left[1+
\frac{{\sf Q}'-{\sf Q}}{{\sf Q}-\rme^{-ip_x}}\right]},
\end{eqnarray}
where we use the notation $Z[{\sf Q}]$ to emphasize the coupling dependence of $Z$.
If ${\sf Q}'$ and ${\sf Q}$ differ only by a finite number of couplings
then the size of the matrix that one needs to take the determinant is finite and given by the rank of ${\sf Q}'-{\sf Q}$.

Let us define ${\sf Q}_0$ as a ${\sf Q}$-matrix for the uniform but possibly non-isotropic system, i.e., $(\V_{j,k},\K_{j,k})=(\V,\K)$ for all $j$ and $k$.
\begin{eqnarray}\label{Quniform}
{\sf Q}_0
=\exp({-\rmi\case{p_y}{2}\sigma_3})\,\rme^{2\D\sigma_2}\,
\exp({\rmi\case{p_y}{2}\sigma_3})\,\rme^{-2\V\sigma_2},
\end{eqnarray} 
where $\theta$ is the dual coupling of $\K$.
We can decompose ${\sf Q}_0$ as
\begin{eqnarray}\label{decomposition}
{\sf Q}_0=\lambda\,\Pi+\lambda^{-1}(1-\Pi),
\end{eqnarray}
where the eigenvalue $\lambda$ and the projection $\Pi$ are written as
\begin{eqnarray}\nn
\lambda&=\xi+\sqrt{\xi^2-1},
\\\label{Pi}
\Pi&=\frac{1}{2}+\rme^{K\sigma_2}\sigma_2\frac{\overline c\,\cos p_y-c-i\sigma_3\,\overline s\,\sin p_y}{2\sqrt{\xi^2-1}}\rme^{-K\sigma_2}\,,
\end{eqnarray}
using the following convenient abbreviations:
\begin{eqnarray}\nn
&(c,s,\overline c,\overline s)=(\cosh2\V,\sinh2\V,\coth2\K,
{\rm csch\,}2\K),
\\\nn
&\xi= c\,\overline c-s\,\overline s\,\cos p_y\,.
\end{eqnarray}
Note that the bar means the horizontal bonds, not complex conjugation.
One can do the same for $\tilde{\sf Q}$ (\ref{Qtilde}) and define $\tilde{\sf Q}_0$ and $\tilde{\Pi}$ in a natural way. $\lambda$ remains the same.

The decomposition (\ref{decomposition}) is useful when one evaluates, for instance, a matrix inversion $({\sf Q}_0-\rme^{-\rmi p_x})^{-1}$.
\begin{eqnarray}\nn
\left[\frac{1}{{\sf Q}_0-\rme^{-ip_x}}\right]_{j,j+m}&=&
\int_0^{2\pi}\frac{dp_x}{2\pi}\rme^{-\rmi mp_x}\left(
    \frac{1-\Pi}{\lambda^{-1}-\rme^{-ip_x}}
    +\frac{\Pi}{\lambda-\rme^{-ip_x}}\right)
\\\nn
&=&\cases{~\lambda^{-|m|-1}\Pi \qquad\qquad~\,\mbox{for}~ m\leq 0
\\ -\lambda^{-m+1}(1-\Pi) \qquad\mbox{for}~ m>0}.
\end{eqnarray}
For $m=0$ the above becomes
\begin{eqnarray}\label{jj}
\left[({\sf Q}_0-\rme^{-\rmi p_x})^{-1}\right]_{jj}=\lambda^{-1}\Pi,
\end{eqnarray}
which will be useful shortly.

Similarly to adding an operator,
one can also add a boundary to the system by changing the couplings in a specific way.
Let us add a boundary vertically at $j=0$, producing two half-planes on both sides of the boundary.
For examples, setting $\K_{0,k}=0$ for all $k$ will impose a free boundary condition, and setting $\K_{0,k}=\infty$ will impose a fixed boundary condition.
We define ${\sf Q}_{\rmb}$ as the corresponding ${\sf Q}$-matrix. 
By definition the difference ${\sf Q}_{\rmb}-{\sf Q}_0$ has non-zero components only at the diagonal block of $j=0$.

To evaluate $\langle{\cal O}\rangle_{\rmb}$, the expectation value of ${\cal O}$ in the presence of the boundary,  
we define ${\sf Q}'$ by implementing the insertion of ${\cal O}$ into ${\sf Q}_{\rmb}$.
Then, $\langle{\cal O}\rangle_{\rmb}$ is written as the following ratio
\begin{eqnarray}\fl\label{ratio}
\langle{\cal O}\rangle_{\rmb}\sim
\frac{Z[{\sf Q}']}{Z[{\sf Q}_{\rmb}]}
    =\sqrt{\prod_{j,k}
    \frac{\sinh\,2\K'_{j,k}}{\sinh\,2\K^{\rmb}_{j,k}}~\det\left[
    \frac{1+({\sf Q}_0-\rme^{-\rmi p_x})^{-1}({\sf Q}'-{\sf Q}_0)}
	{1+({\sf Q}_0-\rme^{-\rmi p_x})^{-1}({\sf Q}_{\rmb}-{\sf Q}_0)}\right]},
\end{eqnarray}
where $\K^{\rmb}$ stands for the horizontal couplings as appeared in ${\sf Q}_{\rmb}$, and $\K'$ stands for those in ${\sf Q}'$. 

To simplify the above equation let us define a {\it Laurent matrix} (see Appendix for the definition) $\Delta_{\rmb}$ and a projection $\eta_x$ as to satisfy
\begin{eqnarray}\nn 
{\sf Q}_{\rmb}-{\sf Q}_0=\Delta_{\rmb}\,\eta_x
\quad\mbox{and}\quad
[\eta_x]_{j,k,\tau,j',k',\tau'}=
\cases{
\delta_{j,j'}\delta_{k,k'}\delta_{\tau,\tau'}\qquad\mbox{for}~j=0
\\
0\qquad\qquad\qquad\mbox{otherwise}}.
\end{eqnarray}
The determinant in (\ref{ratio}) may be rewritten as
\begin{eqnarray}\label{12}
\det\left[1+\frac{1}
    {1+({\sf Q}_0-\rme^{-\rmi p_x})^{-1}\,\Delta_{\rmb}\,\eta_{x}}
    \,\frac{1}{{\sf Q}_0-\rme^{-\rmi p_x}}\,({\sf Q}'-{\sf Q}_{\rmb})\right].
\end{eqnarray}
Using (\ref{jj}) we obtain the identity
\begin{eqnarray}\nn
\frac{1}{1+({\sf Q}_0-\rme^{-\rmi p_x})^{-1}\,\Delta_{\rmb}\,\eta_{x}}
=1-({\sf Q}_0-\rme^{-\rmi p_x})^{-1}\,\Delta_{\rmb}\,\eta_{x}\,\frac{1}{1+\lambda^{-1}\,\Pi\,\Delta_{\rmb}}\,\,,
\end{eqnarray}
and the determinant (\ref{12}) becomes
\begin{eqnarray}\nn
\det\left[1+\left(1-\frac{1}{{\sf Q}_0-\rme^{\rmi p_x}}\,\Delta_{\rmb}
    \,\eta_x\,
    \frac{1}{1+\lambda^{-1}\,\Pi\,\Delta_{\rmb}}\right)
    \,\frac{1}{{\sf Q}_0-\rme^{-\rmi p_x}}\,({\sf Q}'-{\sf Q}_{\rmb})\right].
\end{eqnarray}
Now defining a scalar $\lambda_{\rmb}$ as to satisfy
\begin{eqnarray}\nn
(\lambda_{b}-\lambda)\,\Pi=\Pi\,\Delta_{\rmb}\,\Pi,
\end{eqnarray}
the determinant in (\ref{ratio}) finally becomes
\begin{eqnarray}\fl
\label{longdeterminant}
\det\left[1+\frac{1}{{\sf Q}_0-\rme^{-\rmi p_x}}
\left(1-\Delta_{\rmb}\,\eta_x\,
    \left(1-\lambda_{\rmb}^{-1}\Pi\Delta_{\rmb}\right)
    \frac{1}{{\sf Q}_0-\rme^{-\rmi p_x}}\right)
({\sf Q}'-{\sf Q}_{\rmb})\right].
\end{eqnarray}
Without the boundary, i.e., $\Delta_{\rmb}=0$ and ${\sf Q}_{\rmb}={\sf Q}_0$, we get back the equation (\ref{partitionbulk}). 

The above expression (\ref{longdeterminant}) simplifies dramatically if the operator ${\cal O}$ is located exactly on the boundary and if, at the same time,
one can find a Laurant matrix $\Delta'_{\rmb}$ satisfying
\begin{eqnarray}\nn
{\sf Q}'-{\sf Q}_{\rmb}=\Delta'_{\rmb}\,\eta,
\end{eqnarray}
with a projection $\eta$ that maps into some subspace of $j=0$.\footnote{This is possible for most of boundary correlators but may not be so, for instance, for four point functions of mixed operators.}
The equation (\ref{ratio}) then becomes simply
\begin{eqnarray}\label{shortdeterminant}
\frac{Z[{\sf Q}']}{Z[{\sf Q}_{\rmb}]}
    =\sqrt{\prod_{j,k}
    \frac{\sinh\,2\K'_{j,k}}{\sinh\,2\K^{b}_{j,k}}~
\det\left[1+\frac{\Pi}{\lambda_{\rmb}}\,\Delta'_{\rmb}\,\eta\right]}.
\end{eqnarray}
A similar result is obtained for $\tilde{\sf Q}$ by simply putting
$\tilde{}$ on every matrix and reversing the order of matrix multiplication.

\subsection{Boundary operators in CFT}

We briefly summarize the CFT results on the boundary correlation functions of bcc operators appearing in the Ising model.
At the critical temperature the continuum limit of the Ising model is described by the rational CFT of central charge $\frac{1}{2}$.
There are three primary fields: the identity, the energy, and the spin operator with (holomorphic) dimensions $0,\frac{1}{2}$ and $\frac{1}{16}$ respectively. 
And there are three conformally invariant boundary conditions of which the two are identified with the two fixed boundary states, and the other is identified with the free boundary state.

We summarize the dimensions of bcc operators and their correlation functions \cite{yellowbook}.
\begin{eqnarray}\nn
&\mbox{fixed-to-fixed : }\quad h_{+-}=\case{1}{2}\qquad\, \langle\phi\phi\rangle\sim{|x_1-x_2|}^{-1},
\\\label{CFTprediction}
&\mbox{fixed-to-free\,\,\, : }\quad h_{\pm f}=\case{1}{16}\qquad
\langle\phi\phi\rangle\sim|x_1-x_2|^{-\frac{1}{8}},
\end{eqnarray}
where $x_1$ and $x_2$ are the points on the real axis (boundary) where the bcc operators are located. 

Let us remark on a subtlety when performing a lattice calculation of these correlation functions.
Usually a correlation function $\langle\phi\phi\rangle$ is determined by the ratio $Z'/Z$ where $Z'$ differs from $Z$ by an insertion of $\phi\phi$.
When calculating a correlation function such as the spin-spin correlation function we separate the effect of $\phi\phi$ from that of the background (the bulk and the boundary) by taking the ratio $Z'/Z$ and thus cancelling out the bulk and the boundary free energy.
Without proper cancellation the correlation function will contain a contribution that behaves exponentially with the size of the system.

For the boundary correlation function of bcc operators $\langle\phi\phi\rangle$, however, the full cancellation of the background may not occur since the added operators can {\it change} the boundary state and thereby alter the boundary free energy.
In this case the simple ratio $Z'/Z$ will have an exponential behavior coming from the difference of the boundary free energies in $Z'$ and $Z$.
To extract the correct correlation function we must divide away the exponential factor that comes from the boundary free energy difference.

\section{Boundary correlation function of fixed-to-free bcc operators}

  \begin{figure}
    \centering
    \includegraphics[width=8cm]{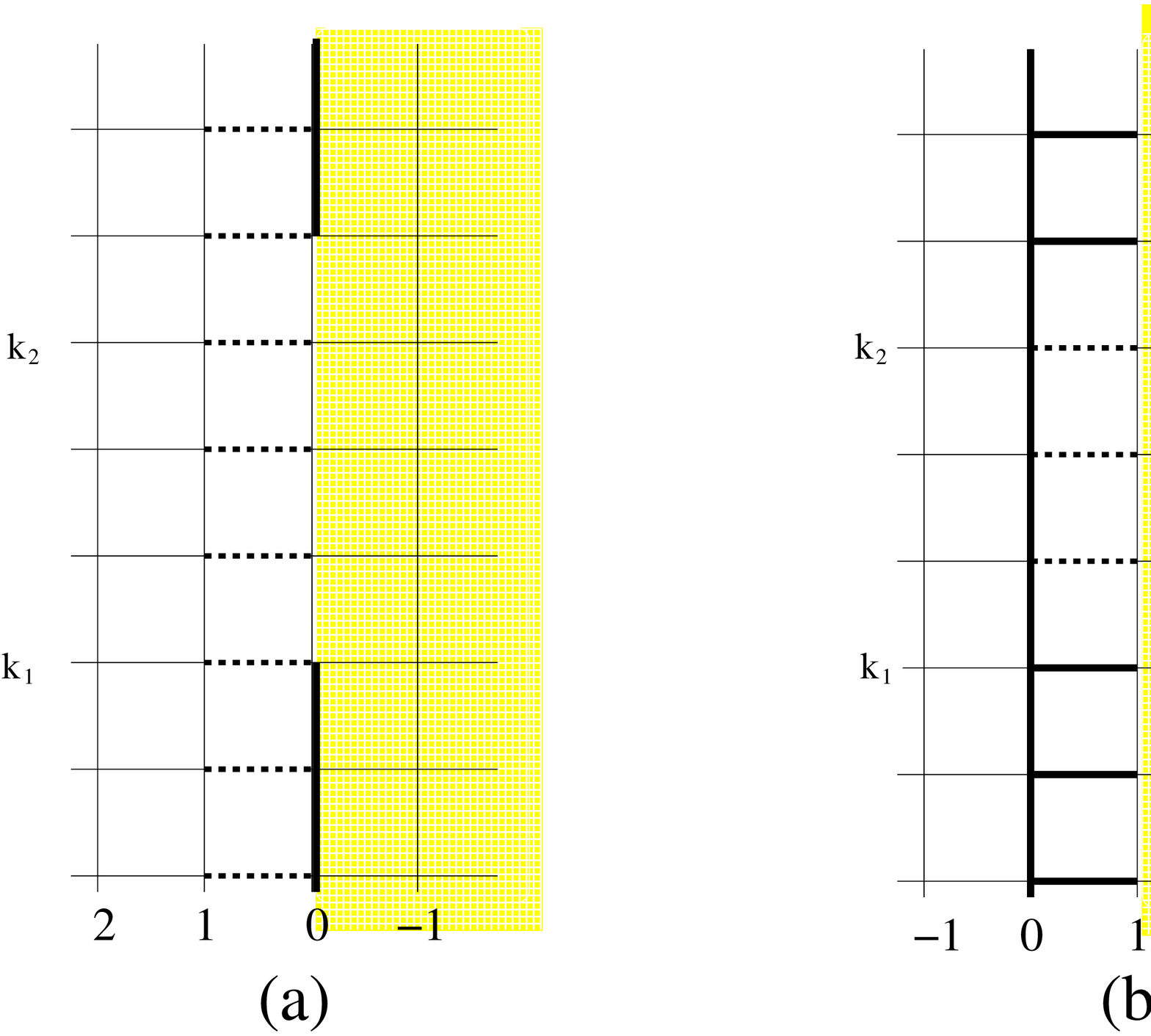}
    \caption{Two lattice realizations of $\langle\mbox{fixed}|\mbox{free}|\mbox{fixed}\rangle$.  The physical side of the half-plane is shadowed (in yellow).  The thick bond is for infinitely strong coupling and the dashed bond is for vanishing coupling. The numbers below indicate $j$ coordinates.  A free boundary is inserted (a) by altering the vertical bonds and (b) by altering the horizontal bonds.  
    Both realizations result in the same configurations in the physical (shadowed) region and, therefore, the same correlation functions. 
    The length of the inserted boundary (free boundary) is $n=k_2-k_1+1$ which, in the case illustrated, equals four.
  \label{IsingBoundary}}
  \end{figure}

Let us proceed to calculate the boundary correlation function of fixed-to-free bcc operators.
By placing two such operators there are two possible situations: a finite free section inside a fixed boundary, and a finite fixed section inside a free boundary.
We will use the notations
$\langle\mbox{fixed}|\mbox{free}|\mbox{fixed}\rangle$ and $\langle\mbox{free}|\mbox{fixed}|\mbox{free}\rangle$ to represent the expectation values for each case.
We start with the former.

Our starting point is equation (\ref{shortdeterminant}). From now on we will simply write $p$ instead of $p_y$, suppressing the subscript.

We assign our lattice uniform but possibly anisotropic couplings: $\V$ for vertical and $\K$ for horizontal bonds. We also assign another coupling $K_\rmf$ to the free boundary.

\subsection{$\langle\mbox{fixed}|\mbox{free}|\mbox{fixed}\rangle$}
There are two ways to realize $\langle\mbox{fixed}|\mbox{free}|\mbox{fixed}\rangle$ on the lattice: one is to insert the free boundary by changing vertical couplings as in figure 1(a) and the other is by changing horizontal couplings as in figure 1(b). We first present the former.

We already defined ${\sf Q}_0$ for uniform couplings in (\ref{Quniform}). 
The fixed boundary may be imposed by setting
\begin{eqnarray}\label{secondway}
\K_{0,k}=\epsilon\quad\mbox{and}\quad\V_{0,k}=-\ln\tilde\epsilon,
\end{eqnarray}
and by taking the limit $\epsilon,\tilde\epsilon\rightarrow0$.
Let us define ${\sf Q}_{\rmb}$ as the corresponding ${\sf Q}$-matrix.

A finite free section is inserted by changing the vertical couplings as
\begin{eqnarray}\label{22}
\V_{0,k}=K_\rmf\quad\mbox{for}\quad k_1\leq k\leq k_2\,,
\end{eqnarray}
 and ${\sf Q}'$ is defined accordingly by implementing (\ref{22}) into ${\sf Q}_{\rmb}$. We set $k_2-k_1+1=n$ so that the number of the inserted free bonds is $n$.

We define the correlation function using (\ref{shortdeterminant}) as
\begin{eqnarray}\nn\fl
\langle\mbox{fixed}|\mbox{free}|\mbox{fixed}\rangle_n&=
{\lim}^*
\left[
(\tilde\epsilon\,\rme^{K_\rmf})^{-n}\,\frac{Z[{\sf Q}']}{Z[{\sf Q}_{\rmb}]}\right]
=
{\lim}^*
\sqrt{(\tilde\epsilon\,\rme^{K_\rmf})^{-2n}\,
{\rm det}\left[1
+\frac{\Pi}{\lambda_{\rmb}}\,\Delta'_{\rmb}\,\eta\right]}
\\\nn&=
{\lim}^*
\sqrt{(\tilde\epsilon\,\rme^{K_\rmf})^{-2n}\,
D_n\left[1+\lambda_{\rmb}^{-1}\Pi{\Delta'_{\rmb}}\right]}.
\end{eqnarray}
${\lim}^*$ stands for the limit $\epsilon,\tilde\epsilon\rightarrow0$.
$D_n$ means the truncated Toeplitz determinant of size $n$ (see Appendix).
In the last equality we used the fact that the rank of the projection $\eta$ is $2n$ which comes from the number of the modified couplings when inserting the free boundary ($2$ comes from the $\tau$-space).

The additional factor of $(\tilde\epsilon\,\rme^{K_\rmf})^{-n}$ comes from the following reason.
The modified $n$ vertical bonds (\ref{22}) have the couplings $K_{\rmf}$ in $Z[{\sf Q}']$ and $(-\ln\tilde\epsilon)$ in $Z[{\sf Q}_\rmb]$; those bonds are weighted differently in the two partition functions.
The additional factor is added to cancel this effect and compare the two configurations (one with the inserted free boundary and the other without) within the same set of couplings $K_\rmf$.

To evaluate the determinant we only need to know the leading terms of $\lambda_{\rmb}$ and $\Delta_{\rmb}'$ in small ($\epsilon,\tilde\epsilon$)-expansions.
They are given by
\begin{eqnarray}\nn\fl
\lambda_\rmb&\approx\frac{1}{4\,\epsilon\,\tilde\epsilon^2}\left(1-\cos p+\frac{(\cos p-1)[(\overline s+\overline s\,\cos p-\overline c)\,s-c]}{\sqrt{\xi^2-1}}\right)
\\\nn\fl
\Delta'_{\rmb}&\approx\frac{1}{4\,\epsilon\,\tilde\epsilon^2}\,
(\sigma_1\,\sin p+\cos p-1)(1-\sigma_2)
\\\nn\fl&\qquad
+\frac{1}{2\,\epsilon}[\cosh2K_\rmf(1+\sigma_2\,\cos p-\sigma_1\,\sin p)-\sinh2K_\rmf(\sigma_2+\cos p-i\sigma_3\,\sin p)].
\end{eqnarray}
The squared correlation function becomes
\begin{eqnarray}\nn\fl
&\langle\mbox{fixed}|\mbox{free}|\mbox{fixed}\rangle_n^2
\\\nn\fl
&={\lim}^*
\left(\tilde\epsilon\,\rme^{K_\rmf}\right)^{-2n}\,D_n
\Bigg[\frac{1+\sigma_2}{2}\left(1+\sigma_1\frac{c\,\overline s\,\cos p-\overline c\,s+\sqrt{\xi^2-1}}{\overline s\,(c-s)\,\sin p}\right)
\\\nn\fl
&\qquad\quad
+2\,\tilde\epsilon^2\,\Pi\,\frac{\cosh2K_\rmf(1+\sigma_2\,\cos p-\sigma_1\,\sin p)-\sinh2K_\rmf(\sigma_2+\cos p-i\sigma_3\,\sin p)}
{1-\cos p+(\xi^2-1)^{-1/2}(\cos p-1)[(\overline s+\overline s\,\cos p-\overline c)\,s-c]}
\Bigg]
\\\nn\fl
&=D_n
\left[\frac{1+\sigma_2}{2}\left(1+\sigma_1\frac{c\,\overline s\,\cos p-\overline c\,s+\sqrt{\xi^2-1}}{\overline s\,(c-s)\,\sin p}\right)
+\frac{1-\sigma_2}{2}\,\left(1-\sigma_1\frac{\cos p+1}{\rme^{4K_\rmf}\sin p}\right)
\right].
\end{eqnarray}
At the last equality we applied the projection $(1-\sigma_2)/2$ to the term of the order $\Or(\tilde\epsilon^2)$ since its $(1+\sigma_2)/2$-projected complement will not contribute to the determinant in our limit.
This procedure may be illustrated using the following simplified equation.
\begin{eqnarray}\label{ab}\fl
\det\left[\left(\begin{array}{cc}\epsilon^{-1}&0\\0&0\end{array}\right)
\left(\begin{array}{cc}a_{11}&a_{12}\\a_{21}&a_{22}\end{array}\right)
+\left(\begin{array}{cc}\not{b_{11}}&\not{b_{12}\,}\\b_{21}&b_{22}\end{array}\right)\right]
\approx
\epsilon^{-1}(a_{11}b_{22}-a_{12}b_{21}),
\end{eqnarray}
where the crossed-out terms, $b_{11}$ and $b_{12}$, do not contribute in the leading order, and
the ``crossing out" is equivalent to applying a projection 
$\tiny\left(\begin{array}{cc}0&0\\0&1\end{array}\right)$
to the $b_{ij}$-matrix.  This procedure will be useful again.

After a simple unitary transformation we obtain the correlation function as
\begin{eqnarray}\nn
\langle\mbox{fixed}|\mbox{free}|\mbox{fixed}\rangle=\sqrt{D_{n}(a^*)}\,,
\end{eqnarray}
where the symbol (see Appendix) $a^*$ is given by 
\begin{eqnarray}\label{astar}
a^*(\rme^{\rmi p})=
\left(\begin{array}{cc}
1&\frac{c\,\overline s\,\cos p-\overline c\,s+\sqrt{\xi^2-1}}{\overline s\,(c-s)\,\sin p}\\-\rme^{-4K_{\rmf}}\cot\frac{p}{2}&1
\end{array}\right).
\end{eqnarray}

Let us present the other way of calculating the same correlation function: by changing horizontal couplings to insert free boundary as illustrated in \ref{IsingBoundary}(b).
Here it is convenient to use the $\tilde{\sf Q}$ notation defined in (\ref{Qtilde}).

First we define $\tilde{\sf Q}_0$  by assuming uniform couplings. 
We add fixed boundary at $j=0$ by assigning 
\begin{eqnarray}\label{firstway}
\K_{0,k}=\V_{0,k}=-\ln\tilde\epsilon,
\end{eqnarray}
and taking the limit $\tilde\epsilon\rightarrow0$.
We define $\tilde{\sf Q}_{\rmb}$ as the corresponding $\tilde{\sf Q}$-matrix.

A finite free section is inserted by cutting off the horizontal bonds as in figure \ref{IsingBoundary}(b)
\begin{eqnarray}\label{25}
\K_{0,k}=\epsilon\quad\mbox{for}\quad k_1<k\leq k_2,
\end{eqnarray}
and taking the limit of $\epsilon\rightarrow0$. Remember that we just changed $n-1$ horizontal couplings (we set $k_2-k_1=n-1$) to insert $n$ free bonds. We define $\tilde{\sf Q}'$ by adding these changes into $\tilde{\sf Q}_{\rmb}$.
In contrast to the previous setup the above procedure restricts the coupling of the free boundary to be $K_\rmf=K$.

Using (\ref{shortdeterminant}) the ratio of the partition functions is written as\footnote{The tilde $\tilde{}$ will be suppressed for the rest of this section though it is implicit in every matrix}
\begin{eqnarray}\nn
\frac{Z[{\sf Q}']}{Z[{\sf Q}_{\rmb}]}=\sqrt{\left(\frac{-\sinh2\epsilon}{\sinh(2\ln\tilde\epsilon)}\right)^{n-1}{\rm det}\left[1
+\eta\,\Delta'_{\rmb}\,\frac{\Pi}{\lambda_{\rmb}}
\right]},
\end{eqnarray}
where the projection $\eta$ is into $k_1<k\leq k_2$ and has the rank $2(n-1)$.

The modified $n-1$ horizontal bonds (\ref{25}) have the infinitely strong
couplings $(-\ln\tilde\epsilon)$ in $Z[{\sf Q}_{\rmb}]$ and vanishingly weak couplings $\epsilon$ in $Z[{\sf Q}']$; those bonds are not weighted equally in $Z[{\sf Q}_{\rmb}]$ and in $Z[{\sf Q}']$. 
We multiply by an additional factor to remedy this effect
and define the correlation function as   
\begin{eqnarray}\label{detfxfrfx}
\langle\mbox{fixed}|\mbox{free}|\mbox{fixed}\rangle_n &=
{\lim}^*\left[
(\tilde\epsilon\,\rme^{\epsilon})^{1-n}
\times\frac{Z[{\sf Q}']}{Z[{\sf Q}_{\rmb}]}\right]
\\\nn&=
{\lim}^*
\sqrt{(4\epsilon)^{n-1}\,
D_{n-1}\left[1
+\Delta'_{\rmb}\,\frac{\Pi}{\lambda_{\rmb}}\right]}.
\end{eqnarray}
${\lim}^*$ stands for the limit $\epsilon,\tilde\epsilon\rightarrow 0$.
The determinant is over the truncated matrix of size $n-1$ which comes from the rank of the projection $\eta$.
 
To evaluate the determinant we only need to know the leading terms of $\lambda_{\rmb}$ and $\Delta_{\rmb}'$ in small ($\epsilon,\tilde\epsilon$)-expansions.
\begin{eqnarray}\nn
\lambda_{\rmb}&\approx&\frac{1}{2\,\tilde\epsilon^2}\left(1+\frac{s\,\overline c-\overline s\,c\,\cos p}{\sqrt{\xi^2-1}}\right),
\\\nn
\Delta'_{\rmb}&\approx&\frac{1}{2\,\epsilon\,\tilde\epsilon^2}\,\frac{1-\sigma_2}{2}\,
\frac{1+\sigma_2\,\cos p-\sigma_1\,\sin p}{2}
-\frac{1}{\tilde\epsilon^2}\frac{1+\sigma_2\,\cos p-\sigma_1\,\sin p}{2}\,.
\end{eqnarray}
The leading term of $\Delta'_{\rmb}$ contains a projection $(1-\sigma_2)/2$ 
and we included the subleading term to evaluate the limit (otherwise the determinant will be zero).   Performing the limiting process by substituting with these leading terms the squared correlation function (\ref{detfxfrfx}) becomes
\begin{eqnarray}\nn\fl
&\langle\mbox{fixed}|\mbox{free}|\mbox{fixed}\rangle_n^2
\\\nn\fl
&={\lim}^*
(4\epsilon)^{n-1}D_{n-1}\left[\textstyle\epsilon^{-1}\,
\frac{1-\sigma_2}{2}\,\frac{1+\sigma_2\,\cos p-\sigma_1\,\sin p}{2}\,\frac{\Pi}{\lambda_{\rmb}}+
\left(1-\frac{1+\sigma_2\,\cos p-\sigma_1\,\sin p}{2}\,\frac{\Pi}{\lambda_{\rmb}}\right)\right]
\\\nn\fl
&=D_{n-1}\left[\textstyle\frac{1-\sigma_2}{2}\,\frac{2(1+\sigma_2\,\cos p-\sigma_1\,\sin p)\,\Pi}{1+(\xi^2-1)^{-1/2}(s\,\overline c-\overline s\,c\,\cos p)}+
\frac{1+\sigma_2}{2}\,\left(2-\frac{2(1+\sigma_2\,\cos p-\sigma_1\,\sin p)\,\Pi}{1+(\xi^2-1)^{-1/2}(s\,\overline c-\overline s\,c\,\cos p)}\right)\right]
\\\nn\fl
&=D_{n-1}\left[\textstyle1+\sigma_2-\sigma_2\, 
\frac{2(1+\sigma_2\,\cos p-\sigma_1\,\sin p)\,\Pi}{1+(\xi^2-1)^{-1/2}(s\,\overline c-\overline s\,c\,\cos p)}\right].
\end{eqnarray}
In the second equality we multiplied the projection $(1+\sigma_2)/2$ to the subleading term as its $(1-\sigma_2)/2$-projected complement will not contribute to the determinant in the limit (see the explanation near the equation (\ref{ab})).

We obtain the correlation function as
\begin{eqnarray}\nn
\langle\mbox{fixed}|\mbox{free}|\mbox{fixed}\rangle_n=\sqrt{D_{n-1}(a)},
\end{eqnarray}
where the symbol $a$ is given by
\begin{eqnarray}\label{a}
a(\rm\rme^{\rmi p})=1+\sigma_2-\sigma_2\, 
\frac{2(1+\sigma_2\,\cos p-\sigma_1\,\sin p)\,\Pi}{1+(\xi^2-1)^{-1/2}(s\,\overline c-\overline s\,c\,\cos p)}\,.
\end{eqnarray}

\subsection{$\langle\mbox{free}|\mbox{fixed}|\mbox{free}\rangle$}

  \begin{figure}
    \centering
    \includegraphics[width=8cm]{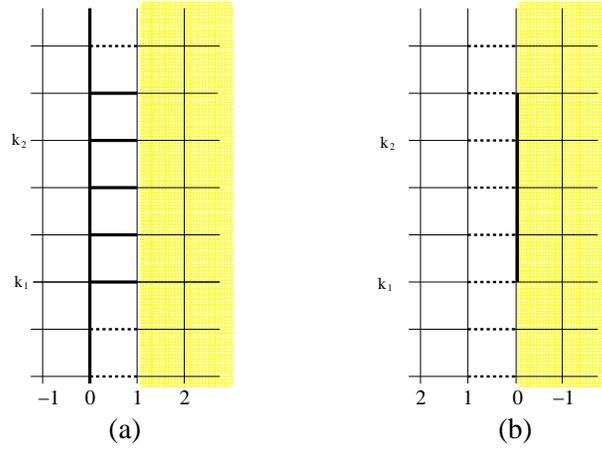}
    \caption{Two lattice realizations of $\langle\mbox{free}|\mbox{fixed}|\mbox{free}\rangle$.  The physical side of the half-plane is shadowed (in yellow).  The thick bond is for infinitely strong coupling and the dashed is bond for vanishing coupling. The numbers below are $j$ coordinates. A fixed boundary is inserted (a) by altering the horizontal bonds and (b) by altering the vertical bonds.   
    Both realizations result in the same configurations in the physical (shadowed) region and, therefore, the same correlation functions. 
    The length of the inserted boundary (free boundary) is $n=k_2-k_1+1$ which, in the case illustrated above, equals four.
  \label{IsingBoundary1}}
  \end{figure}

Next we calculate the reversed boundary state $\langle\mbox{free}|\mbox{fixed}|\mbox{free}\rangle$, where a finite fixed section is inserted into the otherwise free boundary.
Since the procedure is very similar to the previous section we will only describe the setup and the results.

In figure \ref{IsingBoundary1}(a) we insert the fixed boundary by changing horizontal couplings. We define 
$\tilde {\sf Q}_{\rmb}$ to incorporate the free boundary as
\begin{eqnarray}\nn
\K_{0,k}=\epsilon\quad\mbox{and}\quad\V_{0,k}\rightarrow -\ln\tilde\epsilon\,.
\end{eqnarray}
$\tilde{\sf Q}'$ is defined to incorporate the inserted fixed boundary by changing horizontal couplings as
\begin{eqnarray}\nn
\K_{0,k}= -\ln\tilde\epsilon\quad\mbox{for}\quad k_1\leq k\leq k_2+1\,.
\end{eqnarray}
Note that the number of modified bonds is $n+1$ and the length of the inserted fixed boundary is $n=k_2-k_1+1$.
Using (\ref{shortdeterminant}) the correlation function is defined with the appropriate normalization as
\begin{eqnarray}\nn\fl
\langle\mbox{free}|\mbox{fixed}|\mbox{free}\rangle_n={\lim}^*
\sqrt{\left(\tilde\epsilon\,\rme^{\epsilon}\right)^{n+1}\frac{Z[\tilde{\sf Q}']}{Z[\tilde{\sf Q}_{\rmb}]}}
={\lim}^*\sqrt{(4\epsilon)^{-n-1}{D}_{n+1}\left[1
+\tilde\Delta'_{\rmb}\,\frac{\tilde\Pi}{\tilde\lambda_{\rmb}}
\right]}.
\end{eqnarray}
After similar steps as in section 3.1
we get
\begin{eqnarray}\nn
\langle\mbox{free}|\mbox{fixed}|\mbox{free}\rangle_n=\sqrt{D_{n+1}(b)},
\end{eqnarray}
where the symbol $b$ is given by
\begin{eqnarray}\label{bp}
b(\rme^{\rmi p})=\frac{1}{2}
\left(\begin{array}{cc}
1&-\cot\frac{p}{2}\\\frac{\overline c\,s\,\cos p-\overline s\,c+\sqrt{\xi^2-1}}{s\,(\overline c-\overline s)\,\sin p}&1
\end{array}\right).
\end{eqnarray}

In figure \ref{IsingBoundary1}(b) we calculate the same quantity by changing the vertical bonds to insert the fixed boundary. We define ${\sf Q}_{\rmb}$  to incorporate the free boundary as
\begin{eqnarray}\nn
\K_{0,k}=\epsilon\quad\mbox{and}\quad\V_{0,k}\rightarrow K_{\rmf}\,.
\end{eqnarray} 
We define ${\sf Q}'$ to incorporate the inserted fixed boundary as
\begin{eqnarray}\nn
\V_{0,k}=-\ln\tilde\epsilon\quad\mbox{for}\quad k_1\leq k\leq k_2\,.
\end{eqnarray}
With an appropriate normalization factor the correlation function is obtained as
\begin{eqnarray}\nn\fl
\langle\mbox{free}|\mbox{fixed}|\mbox{free}\rangle_n
=
{\lim}^*\sqrt{\left(\tilde\epsilon\,\rme^{K_{\rmf}}\right)^{n}\frac{Z[{\sf Q}']}{Z[{\sf Q}_{\rmb}]}}
={\lim}^*\sqrt{\left(\tilde\epsilon\,\rme^{K_{\rmf}}\right)^{n}{D}_{n}\left[1
+\frac{\Pi}{\lambda_{\rmb}}\,\Delta'_{\rmb}\right]}.
\end{eqnarray}
After similar steps to section 3.1
we get
\begin{eqnarray}\nn
\langle\mbox{free}|\mbox{fixed}|\mbox{free}\rangle_n=\sqrt{D_{n}(b^*)},
\end{eqnarray}
where the symbol $b^*(\rme^{\rmi p})$ is given in a rather complicated form by
\begin{eqnarray}\label{b*}\fl
b^*(\rme^{\rmi p})=\frac{1+\sigma_2}{2}+\frac{\rme^{-K_\rmf}(\bar s+\bar c)}{2s\sin^2 p}\times
\\\nn\fl\qquad
\textstyle
\frac{-\sqrt{\xi^2-1}(\cosh2K_\rmf-\sinh2K_\rmf\cos p)+\cosh2K_\rmf(c\bar s-s\bar c\cos p)+\sinh2K_\rmf(-\bar sc\cos p+s\bar c-\bar ss\sin^2 p)}{\bar sc\cos p-s\bar c+\cosh 4K_\rmf(-\bar sc\cos p-\bar ss)+\sinh4K_\rmf(\bar sc+\bar ss\cos p)}
\\\nn\fl\qquad\times
(-\sigma_2c\bar s-\sigma_1s\bar s\sin p+s\bar c)(1+\sigma_2\cos p+\sigma_1\sin p)(\cosh K_\rmf-\sigma_2\sinh K_\rmf).
\end{eqnarray}

\section{Some identities and the Kramers-Wannier duality}

The symbols: $a^*$ (\ref{astar}), $a$ (\ref{a}), $b$ (\ref{bp}) and $b^*$ (\ref{b*}), that we have obtained in the previous section all depend on $K$ and $\K$, and $a^*$ and $b^*$ have further dependence on $K_\rmf$, the coupling of the free boundary.  By setting $K_\rmf=K$ we expect $a^*$ (and $b^*$) to produce the same correlation function as $a$ (and $b$) since the latter has the boundary coupling of $K$. More precisely, we expect
\begin{eqnarray}\label{4det}
D_n(a^*)\Big|_{K_\rmf=K}=D_{n-1}(a),\qquad
D_n(b^*)\Big|_{K_\rmf=K}=D_{n+1}(b).
\end{eqnarray}

Though the second identity is trivial from the physical point of view
the first identity is not totally obvious.  
Its RHS corresponds to figure \ref{IsingBoundary}(b) and, there, the two separated fixed boundaries must be of the same sort since they are actually connected underneath the free boundary through the unphysical side.
However, in LHS which corresponds to figure \ref{IsingBoundary}(a) the two separated fixed boundaries can be polarized in different spin states, i.e., spin-up for one side and spin-down for the other. 
The identity then means that the ``up-down" configurations are of measure zero compared to the ``up-up" or ``down-down" configurations.

This can be understood by considering a one point function of fixed-to-fixed bcc operator and the Kramers-Wannier duality.  We will show that this one point function vanishes, which will mean that the ``up-down" configuration is indeed of measure zero.  The Kramers-Wannier duality says that the lattice with couplings $\{K\}$ has the same partition function as the dual lattice with couplings $\{K^*\}=\{{\rm arctanh\,}{\rm e}^{-2K}\}$. If one changes $K$ into $K+{\rm i}\frac{\pi}{2}$
then its dual coupling changes as $K^*\rightarrow -K^*$, which is equivalent to exchanging the spin operators and the disorder operators \cite{Kadanoff2}.
Let us polarize the spins on the vertical boundary by imposing a positive infinite couplings $K=\infty$ along the boundary.  Then we change the sign of one bond to $K=-\infty$, which is to put a fixed-to-fixed bcc operator.  In the dual lattice, the (vertical) bond with the infinitely strong coupling becomes disconnected (horizontal) bonds (i.e. $K^*=0$) and the (vertical) bond with the coupling $K=-\infty$ becomes the (horizontal) bond with the coupling $K^*={\rm i}\frac{\pi}{2}$, which is simply to put a pair of spin operators on that bond.   
The partition function then vanishes naturally because one point function of the spin operator vanishes. 

We can also understand the similarity of $a^*$ and $b$ from the Kramers-Wannier duality.  The similarity may be explicitly expressed as
\begin{eqnarray}\label{39}
D_n(a^*)\Big|_{K_\rmf=0,\,c\leftrightarrow\overline c,\,s\leftrightarrow\overline s}=2^{2n}D_n(b).
\end{eqnarray} 
By imposing $K_\rmf=0$ we have chosen that the inserted free boundary (four bonds in the figure 1(a)) have vanishing couplings, that is, disconnected bonds.
Since the Kramers-Wannier duality exchanges the fixed bond with the disconnected bond one obtains the figure \ref{IsingBoundary1}(a) as the dual configuration.  For the other bonds, the duality acts as $c\leftrightarrow\overline c$ and $s\leftrightarrow \overline s$.

The above paragraph explain the equation (\ref{39}) except the factor of $2^{2n}$. To explain the factor let us explicitly write the Kramers-Wannier duality as follows.
\begin{eqnarray}
\fl
\sum_{\{s\}}\prod_{(ij)}\rme^{Ks_is_j}&=\prod_{\mbox{\tiny bonds}}\cosh K\sum_{\mbox{\tiny loops}}(\tanh K)^{\mbox{\tiny length}}\qquad\mbox{(FK~loops)}
\\&=2\,\prod_{\mbox{\tiny bonds}^*}\rme^{K}\sum_{\mbox{\tiny loops}^*}\rme^{-2K\mbox{\tiny length}^*}\qquad\mbox{(Ising domain~walls)}.
\end{eqnarray}
For each case the partition function has two parts: the product over all the bonds and the summation over loop ensemble.  Since our Toeplitz determinants are the ratio of two partition functions we may express them using the above form.  If we express $D_n(a^*)\big|_{K_\rmf=0,\cdots}$ by FK loops and $D_n(b)$ by the Ising domain walls then the contributions from the loop-summations are equal; only the contributions from the bond-products may differ.  By noting that $\cosh K_{\infty}\approx\frac{1}{2}\rme^{K_{\infty}}$ for $K_{\infty}\rightarrow\infty$ and also that we normalized the correlation functions by dividing out the infinite factor $\rme^{K_{\infty}}$,  we can confirm the existence of the factor $2^{2n}$.

From the symbol $a^*$ we may extract more information than just the correlation function. If we series-expand $\sqrt{D_n(a^*)}$ in powers of $\rme^{-4K_\rmf}$ then each term collects the configurations with the same number of ``boundary islands" which we will define as the region of oppositely oriented spins within the uniformly fixed boundary. The number of boundary islands is given by the power of $\rme^{-4K_\rmf}$.  

The maximal number of boundary islands corresponds to $\frac{1+(-1)^n}{2}\rme^{-2nK_\rmf}\sqrt{T_n(a^*_{1,2})}$ where $a^*_{1,2}$ is the $(1,2)$-component of the symbol $a^*$.  
This corresponds to the insertion of an alternating boundary within the fixed boundary. Therefore we get the following as a byproduct.
\begin{eqnarray}
\langle\mbox{fixed}|\mbox{alternating}|\mbox{fixed}\rangle_n
=\textstyle\frac{1+(-1)^n}{2}\sqrt{T_n(a^*_{1,2})}.
\end{eqnarray}
The boundary condition must alternate even times for the correlation function to be non-zero. 

\section{Asymptotic behavior of the determinant}

The asymptotic behavior of a general block Toeplitz determinant is obtained by the generalized Szeg\"o's theorem which is proven by Widom \cite{WidomBlock} (see Appendix).  This theorem may not apply, however, when the (matrix) symbol contains some singularities. For a scalar Toeplitz matrix there is the Fisher-Hartwig theorem (see Appendix) which tells you the asymptotic behavior of the determinant when its symbol contains the Fisher-Hartwig singularities.
For block Toeplitz matrices there 
seems to be no analogous theorem, which we will need to study the correlation function at the critical temperature.

Interestingly, however, $D(a^*)$ can be transformed into a scalar Toeplitz determinant.
Let us recall the definition of $a^*$ (\ref{astar}).
\begin{eqnarray}\nn\fl
a^*(\rme^{\rmi p})=\left(\begin{array}{cc}
1&w(p)\\-\rme^{-4K_{\rmf}}\cot\frac{p}{2}&1
\end{array}\right),\qquad \mbox{where}\qquad w(p)=\frac{c\,\overline s\,\cos p-\overline c\,s+\sqrt{\xi^2-1}}{\overline s\,(c-s)\,\sin p}.
\end{eqnarray}
By performing a simple similarity transformation we get 
\begin{eqnarray}\nn\fl
D_n(a^*)&=\det\left(\begin{array}{cc}
I_n&T_n(w)\\-\rme^{-4K_{\rmf}}T_n(\cot\frac{p}{2})&I_n
\end{array}\right)=\det\left(\begin{array}{cc}
I_n&-\rme^{-4K_{\rmf}}T_n(w)T_n(\cot\frac{p}{2})\\I_n&I_n
\end{array}\right)
\\\nn\fl&=\det\Big[I_n+\rme^{-4K_\rmf}T_n(w)T_n(\cot\textstyle\frac{p}{2})\Big].
\end{eqnarray}
A further simplification can be made by using the fact: $\big[T_n(\cot\frac{p}{2})\big]^{-1}=T_n(\tan\frac{p}{2})$ which holds for an even $n$. Multiplying this matrix inside the determinant we get
\begin{eqnarray}\label{scalardet}
D_n(a^*)=(-1)^{n/2}D_n\Big(\tan\textstyle\frac{p}{2}+\rme^{-4K_\rmf}w(p)\Big),\qquad \mbox{for~even~}n.
\end{eqnarray}
We have used $D_n(\cot\frac{p}{2})=(-1)^{n/2}$ for an even $n$. 

Let us find out the asymptotic behavior of the scalar Toeplitz determinant (\ref{scalardet}) of which the symbol is the scalar function,
$$h(\rme^{\rmi p})\equiv \tan\textstyle\frac{p}{2}+\rme^{-4K_\rmf}\frac{c\,\overline s\,\cos p-\overline c\,s+\sqrt{\xi^2-1}}{\overline s\,(c-s)\,\sin p}.$$
According to the Szeg\"o's strong limit theorem (see Appendix) the asymptotic 
behavior of the determinant is given by
\begin{eqnarray}\nn
D_n(h)\sim G(h)^nE(h)\qquad \mbox{as}\qquad n\rightarrow\infty.
\end{eqnarray}
The exponential factor is simply given by
\begin{eqnarray}\nn
G(h)=\exp(\log h)_0=\exp\int_0^{2\pi}\frac{dp}{2\pi}\log h(\rme^{\rmi p}).
\end{eqnarray}
Let us evaluate $(\log h)_0$ assuming $\K=\V$ and $K_\rmf=K$ for simplicity.
\begin{eqnarray}\label{cK}
(\log h)_0=\int_0^{2\pi}\frac{dp}{2\pi}\log\left[2\,\rme^{-2K}\Big(c-c\,s+s-s\,\cos
p+\xi\Big)\right].
\end{eqnarray}
Near the critical temperature, $K\sim K_c=\frac{1}{2}\ln(1+\sqrt2)$, this becomes
\begin{eqnarray}\label{singular}
(\log h)_0=\textstyle \frac{2}{\pi}C_G+\log2(\sqrt2-1)+\frac{4}{\pi}(K-K_c)\log|K-K_c|+\cdots,
\end{eqnarray}
where $C_G\approx0.915966$ is called the Catalan's constant.

This exponential factor, $G(h)=\exp(\log h)_0$, has the following clear physical meaning. Let us prepare the boundary condition such that all the spins to one side of a certain boundary bond are polarized into the up-state. Then the probability that the free spin nearest to the polarized spins will also polarize to up-state, is given by $G(h)^{-1}$.

At the critical temperature, as explained near the end of the section 2.3, this exponential factor indicates the boundary free energy difference.  In our case, $(\log h)_0$ is {\it twice} the boundary free energy difference (per bond)  between the free boundary and the fixed boundary because our correlation functions are the square root of the determinant.

Let us calculate the asymptotic behavior of the determinant $D_n(a^*)$ (\ref{scalardet}) at the critical temperature, $K=K_c=\frac{1}{2}{\rm arcsinh}\,1$. We also set the boundary couplings as $K_\rmb=K_c$. Then we need to consider the following symbol of the scalar Toeplitz matrix.
\begin{eqnarray}
h_c(\rme^{\rmi p})=\textstyle{\frac{\sqrt2-1}{\sqrt2}\sec\frac{p}{2}}\left(\sqrt{1-\cos p}+\sqrt{3-\cos p}\right).
\end{eqnarray}
One notices that this symbol has a jump discontinuity and can be decomposed as
\begin{eqnarray}
h_c(t)=(-t)^{1/2}\,b_c(t),
\end{eqnarray}
where $b_c(t)$ does not have any Fisher-Hartwig type singularity (see Appendix) but contains a simple pole divergence at $t=-1$.
This pole singularity prohibits us from a simple Wiener-Hopf decomposition of the function $\log b_c(t)$ and from using the following formula presented in Appendix.
\begin{eqnarray}\fl
D_n(h_c)\sim\rme^{n(\log h_c)_0}{\textstyle\frac{G(\frac{3}{2})G(\frac{1}{2})}{G(1)}}\exp\Big[\sum_{k=1}^\infty k(\log b_c)_k(\log b_c)_{-k}\Big]b_+(1)^{1/2}b_-(1)^{1/2}n^{\alpha^2-\beta^2}.
\end{eqnarray}
Instead, as we did naively in Appendix, let us try to evaluate the asymptotic behavior by using the Szeg\"o's theorem in a formal way.

\begin{figure}
    \centering
    \includegraphics[width=8cm]{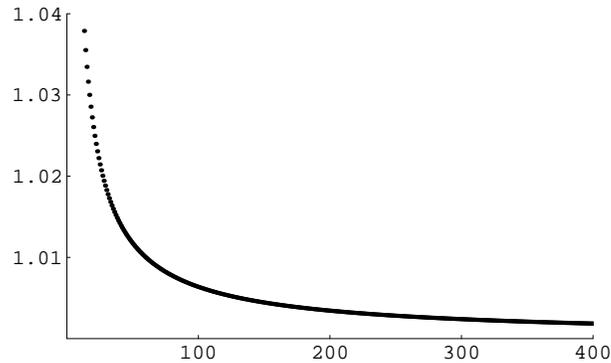}
    \caption{$D_n(h_c)\big/\left[(\frac{2}{\pi}C_G)^n0.5506047747\times n^{-1/4}\right]$ vs. $n$. It shows the convergence to one as $n$ goes to $\infty$.
  \label{final}}
  \end{figure}

Let us consider the ratio $D_n(h_c)/D_n(\sqrt{-t})$.
Its exponential factor, $\rme^{(\log h_c)_0}$, was already obtained to be $\frac{2}{\pi}C_G$ at (\ref{singular}).  The constant factor is written as follows by formally applying the Szeg\"o's theorem.
\begin{eqnarray}\label{55}
\left(\textstyle\frac{2}{\pi}C_G\right)^{-n}{\textstyle\frac{D_n(h_c)}{D_n(\sqrt{-t})}}=\exp\Big[\sum_{k=1}^\infty k(\log h_c)_k(\log h_c)_{-k}+\sum_{k=1}^\infty{\textstyle\frac{1}{4k}}\Big].
\end{eqnarray}
We calculate $(\log h_c)_k$ for $k\neq 0$ using integration by part as
\begin{eqnarray}\nn\fl
(\log h_c)_k&=\int_0^{2\pi}\frac{dp}{2\pi}\rme^{-\rmi k p}\Big(\log\sqrt t-\log(1+\rme^{\rmi p})+\log({\textstyle\sqrt{1-\cos p}+\sqrt{3-\cos p}})\Big)
\\\nn\fl&=-\frac{1}{2k}+\frac{(-1)^k}{k}\delta_{k,|k|}+\frac{\alpha_{|k|}}{|k|},\qquad\mbox{where}\qquad\rmi\alpha_{|k|}{\rm sgn}(k)\equiv{\textstyle\int_0^{2\pi}\frac{dp}{2\pi}\frac{\rme^{-\rmi kp}\cos(p/2)}{\sqrt{6-2\cos p}}}\,.
\end{eqnarray}
In the above equation we considered $\log(1+\rme^{\rmi p})$ as  $\lim_{\epsilon\rightarrow0+}\log(1+\epsilon+\rme^{\rmi p})$ hoping that the final result may not depend on the specific limiting process.  Substituting this into the equation (\ref{55}) we can cancel out the diverging harmonic series and obtain a finite quantity as follows.
\begin{eqnarray}\nn
\left(\textstyle\frac{2}{\pi}C_G\right)^{-n}{\textstyle\frac{D_n(h_c)}{D_n(\sqrt{-t})}}=\exp\Big[\sum_{k=1}^\infty\frac{\alpha_k^2}{k}+\sum_{k=1}^\infty\frac{(-1)^k\alpha_k}{k}+\sum_{k=1}^\infty\frac{(-1)^k}{2k}\Big].
\end{eqnarray}
These sums can be evaluated using the relation $\frac{\cos(p/2)}{\sqrt{6-2\cos p}}=i\sum_{k=-\infty}^\infty\alpha_{|k|}{\rm sgn}(k)\rme^{\rmi kp}$ and its integrated form, ${\rm arctanh}\frac{\sqrt2\sin(p/2)}{\sqrt{3-\cos p}}=\sum_{k\neq0}\frac{\alpha_{|k|}}{|k|}\rme^{\rmi kp}+\frac{2C_G}{\pi}$, as
\begin{eqnarray}\fl\nn
\sum_{k=1}^\infty\frac{\alpha_k^2}{k}=-\int\frac{dp}{2\pi}\int\frac{dp'}{2\pi}\log\Big|2\sin\frac{p-p'}{2}\Big|\frac{\cos(p/2)}{\sqrt{6-2\cos p}}\frac{\cos(p'/2)}{\sqrt{6-2\cos p'}}\approx0.03921085231,
\\\nn\fl
\sum_{k=1}^\infty\frac{(-1)^k\alpha_k}{k}=\frac{1}{2}{\rm arctanh}\frac{1}{\sqrt2}-\frac{C_G}{\pi},\qquad \sum_{k=1}^\infty\frac{(-1)^k}{2k}=-\frac{\log2}{2}.
\end{eqnarray}
Collecting all these we finally obtain the asymptotic behavior of the determinant as
\begin{eqnarray}\fl\nn
D_n(h_c)&\sim \left(\textstyle\frac{2}{\pi}C_G\right)^{n}\exp\Big[{\textstyle0.03921085231+\frac{1}{2}{\rm arctanh}\frac{1}{\sqrt2}-\frac{C_G}{\pi}-\frac{\log2}{2}}\Big]D_n(\sqrt{-t})
\\\nn\fl&\sim\left(\textstyle\frac{2}{\pi}C_G\right)^{n}\exp\Big[{\textstyle0.03921085231+\frac{1}{2}{\rm arctanh}\frac{1}{\sqrt2}-\frac{C_G}{\pi}-\frac{\log2}{2}}\Big]\Big(2^{1/12}\rme^{1/4}A^{-3}\Big)n^{-1/4}
\\\nn\fl&\approx \left(\textstyle\frac{2}{\pi}C_G\right)^{n}0.5506047747\times n^{-1/4},
\end{eqnarray}
where $A=\exp[\frac{1}{12}-\zeta'(-1)]=1.282427129...$ is the Glaisher-Kinkelin constant and $\zeta'$ is the derivative of the Riemann zeta function.  Figure \ref{final} confirms this
 asymptotic behavior of the determinant.

\section{Conclusion}

Starting from the formulation of \cite{Kadanoff} we obtained the general boundary correlation function in a $2\times2$ block Toeplitz determinant (\ref{shortdeterminant}).
We then calculated the boundary correlation function of fixed-to-free bcc operator.
Four lattice realizations of this correlation function lead to four block Toeplitz determinants. 
We explained some interesting properties between them using the Kramers-Wannier duality.
These determinants could be transformed into a scalar determinant when the size of the matrix is even. Using this scalar Toeplitz determinant we calculated the asymptotic behavior of the correlation function at large distance, which agrees with the conformal field theory.

\ack
I would like to thank Leo Kadanoff for his guidance throughout the project. I also thank Ilya Gruzberg and Eldad Bettelheim for useful discussions. This research is supported by ASC-FLASH.

\section*{Appendix. Toeplitz matrix: Szeg\"o's theorem, and Fisher-Hartwig determinants}

Here we collect definitions and theorems about Toeplitz matrices \cite{book}.

Given a sequence $\{a_n\}^\infty_{n=-\infty}$ of complex numbers let us define an infinite matrix $M$ such that its elements are given by $M_{ij}=a_{i-j}$ for $-\infty<i,j<\infty$. Such matrices, that is, doubly-infinite matrices which are constant along the diagonals, are called {\it Laurent matrices}.
Let us assume that there is a function $a(t)$ defined on a unit circle, $\{t|t=\rme^{\rmi p},p\in[0,2\pi)\}$ such that   
\begin{eqnarray}\label{a_n=}
a_n=\frac{1}{2\pi}\int_0^{2\pi}a(\rme^{\rmi p})\,\rme^{-\rmi n p}\,dp.
\end{eqnarray}
The function $a$ is referred to as the {\it symbol} of the matrix $M$.

A {\it Toeplitz matrix} is an infinite sub-matrix of a Laurant matrix such as follows.
\begin{eqnarray}\label{Toeplitz}T(a)=\left(
\begin{array}{ccccc}
a_0	& a_{-1}&a_{-2}	&a_{-3}  & \cdots \\
a_1     & a_0 	& a_{-1}&a_{-2}  & \cdots \\
a_2 	& a_1 	& a_0   &a_{-1}  & \cdots \\
a_3  	&a_2  	&a_1    &a_0     & \cdots \\
\vdots	&\vdots	&\vdots &\vdots  & \ddots
\end{array}\right),
\end{eqnarray}
where the function $a$ is defined by (\ref{a_n=}) and is called the {\it symbol} of the matrix $T(a)$ in this context. 
If the symbol $a$ is a finite dimensional matrix function then the resulting matrix is a {\it block Toeplitz matrix}.

We are mostly concerned with the {\it truncated Toeplitz matrix} of size $n\times n$ which will be denoted by $T_n(a)$. The determinant of $T_n(a)$ will be denoted by $D_n(a)$.
\begin{eqnarray}\nn
D_n(a)=\det T_n(a)=\det\left|\begin{array}{cccc}a_0&a_{-1}&\cdots&a_{1-n}
\\a_{1}&a_0&\cdots&a_{2-n}
\\\vdots&\vdots&\ddots&\vdots
\\a_{n-1}&a_{n-2}&\cdots&a_0
\end{array}\right|.\end{eqnarray}

When the symbol $a$ is sufficiently smooth, non-zero, and satisfies $\int_0^{2\pi}\frac{dp}{2\pi}\frac{d}{dp}\log a(\rme^{\rmi p})=0$ (i.e., zero winding number), 
the {\it Szeg\"o's strong limit theorem} states that the determinant $D_n(a)$ has the asymptotic behavior given by
\begin{eqnarray}\nn
D_n(a)\sim G(a)^nE(a)\qquad n\rightarrow\infty,
\end{eqnarray}
where $G(a)$ and $E(a)$ are defined by
\begin{eqnarray}\nn
G(a)=\exp(\log a)_0,\qquad E(a)=\exp\sum_{k=1}^\infty k(\log a)_k(\log a)_{-k},
\end{eqnarray}
using the fourier components, $(\log a)_n:=\int_0^{2\pi}\frac{dp}{2\pi}\log a(\rme^{\rmi p})$.
This theorem is generalized to block Toeplitz matrices by Widom \cite{WidomBlock}. 

The symbols with {\it pure Fisher-Hartwig singularities} are characterized by two numbers $\alpha$ and $\beta$ as
\begin{eqnarray}\nn
w_{\alpha,\beta}(\rme^{\rmi p})
=|2-2\cos (p-p_0)|^{\alpha}\,\rme^{i\beta(p-\pi)}\,,
\end{eqnarray}
where $\alpha$ deals with the {\it zero-modulus singularity} and $\beta$ deals with the {\it jump singularity}.
For some range of $\alpha$ and $\beta$, the asymptotic behavior of $D(w_{\alpha,\beta})$ is known to be
\begin{eqnarray}\label{D_nw_ab}
D_n(w_{\alpha,\beta})\sim \frac{G(1+\alpha+\beta)G(1+\alpha-\beta)}{G(1+2\alpha)}n^{\alpha^2-\beta^2}\qquad n\rightarrow\infty.
\end{eqnarray}
$G$ is the so-called {\it Barnes G-function} which is an entire function defined by
\begin{eqnarray}
G(z+1)=(2\pi)^{z/2}\rme^{-z(z+1)/2-C_\gamma z^2/2}\prod_{n=1}^\infty\Big\{(1+{\textstyle\frac{z}{n}})^n\,\rme^{-z+z^2/(2n)}\Big\},
\end{eqnarray}
where $C_\gamma\sim0.57721...$ is the Euler-Mascheroni constant.

In general, when the symbol contains certain singularities, 
the sum $\sum_{k=1}^\infty k(\log a)_k(\log a)_{-k}$ appearing in the definition of $E(a)$ may diverge and the determinant may pick up a power-law behavior.
Such symbol may decompose into the product of a finite number of pure Fisher-Hartwig singularities and a ``nice" function (i.e., a function subject to the condition that applies to Szeg\"o's theorem). For an example, let us write
\begin{eqnarray}\label{64}
a(t)=\exp(\log a)_0\,w_{\beta}(t)\,b(t)\qquad w_{\beta}(t)=(-t)^\beta,
\end{eqnarray}
and assume that $b(t)$ is a ``nice" function that does not have any singularity.  Let us calculate the ratio, $\frac{D_n(a)}{D_n(w_{\beta})}$, by formally applying the Szeg\"o's theorem respectively on both quantities as follows.
\begin{eqnarray}\fl\nn
\frac{D_n(a)}{D_n(w_{\beta})}=G(a)^n\exp\Big[{
\sum_{k=1}^\infty k(\log a)_k(\log a)_{-k}-\sum_{k=1}^\infty k(\log w_\beta)_k(\log w_\beta)_{-k}}\Big]
\\\nn\fl=G(a)^n\exp\Big[\sum_{k=1}^\infty k(\log b)_k(\log b)_{-k}
+\sum_{k=1}^\infty k(\log b)_k(\log w_\beta)_{-k}+\sum_{k=1}^\infty k(\log w_\beta)_{k}(\log b)_{-k}\Big],
\end{eqnarray}
where all three terms converge since $b$ is ``nice."
By using $(\log w_\beta)_k=-\frac{\beta}{k}(1-\delta_{k0})$, we can simplify the above as
\begin{eqnarray}
\nn\fl=G(a)^n\exp\Big[\sum_{k=1}^\infty k(\log b)_k(\log b)_{-k}
+\beta\sum_{k=1}^\infty(\log b)_k-\beta\sum_{k=1}^\infty(\log b)_{-k}\Big]
\\\nn\fl=G(a)^n\exp\Big[\sum_{k=1}^\infty k(\log b)_k(\log b)_{-k}\Big]
b_+(1)^\beta\,b_-(1)^{-\beta},
\end{eqnarray}
where $b_{\pm}$ comes from factorization, $b=b_+b_-$, where 
$b_+(t)=\exp[\sum_{n=0}^\infty(\log b)_nt^n]$ and $b_-(t)=\exp[\sum_{n=-\infty}^{-1}(\log b)_nt^n].$ 
Combined with (\ref{D_nw_ab}) the above equation tells you the asymptotic behavior of the determinant, $D_n(a)$.
This is only a physicist's way of presenting the more general theorem which is proved by Ehrhardt and Silbermann \cite{ES}.

\end{document}